\documentclass[12pt,preprint]{aastex}

\slugcomment{Accepted for publication in ApJ}

\shorttitle{Maser Properties of IRAS~19312+1950}
\shortauthors{Nakashima et al.}

\begin{document}


\title{Maser Properties of the Enigmatic IRAS Source 19312+1950}


\author{Jun-ichi Nakashima\altaffilmark{1,2}, Shuji Deguchi\altaffilmark{3}, Hiroshi Imai\altaffilmark{4},\\ Athol Kemball\altaffilmark{2,5} and B. M. Lewis\altaffilmark{6}}

\altaffiltext{1}{Department of Physics, University of Hong Kong, Pokfulam Rd., Hong Kong; email(JN): junichi@hku.hk}

\altaffiltext{2}{Department of Astronomy, University of Illinois at Urbana-Champaign, 1002 West Green Street, Urbana, IL 61801}

\altaffiltext{3}{Nobeyama Radio Observatory, Minamimaki, Minamisaku, Nagano 384-1305, Japan}

\altaffiltext{4}{Department of Physics and Astronomy, Graduate School of Science and Engineering, Kagoshima University, 1-21-35 Korimoto, Kagoshima 890-0065, Japan}

\altaffiltext{5}{National Center for Supercomputing Applications, University of Illinois at Urbana-Champaign, 605 East Springfield Avenue, Champaign, IL 61820}

\altaffiltext{6}{Arecibo Observatory, HC3, Box 53995, Arecibo, PR 00612}


\begin{abstract}
The IRAS source, 19312+1950, exhibits SiO maser emission, which is predominantly detected in evolved stars enshrouded by a cold molecular envelope. In fact, the mojority of the observational properties of IRAS~19312+1950 is consistent with the nature of an asymptotic giant branch (AGB) star or post-AGB star. Interestingly, however, some of the observational properties cannot be readily explained within the standard scheme of stellar evolution, and those are rather reminiscent of young stellar objects. In the present research we considered the evolutionary status of IRAS~19312+1950 as revealed by the VLBI and MERLIN observations in SiO, H$_2$O and OH maser lines. The double-peaked profile of the 22~GHz H$_2$O maser line is clearly detected, with the emission regions of its red and blue-shifted components separately located, leaving a space of about 10.9~mas between them. The kinematic properties of H$_2$O maser emission region appear to be more consistent with a bipolar flow rather than other interpretations such as the Keplerian rotation of a disk. The red-shifted component of the SiO maser emission, which exhibits a double-peak profile in previous single-dish observations, is clearly detected in the present interferometry, while the 1612~MHz OH maser line exhibits a complicated line profile consisting of a single strong peak and many weak, high-velocity spikes. The structure of OH maser emission region is partially resolved, and the kinematic properties of the OH maser emission region are reminiscent observations of a spherically expanding shell, even though the evidence is scant. Collectively, the maser observations described here provide additional support for the evolved star hypothesis for IRAS~19312+1950.
\end{abstract}




\section{Introduction}

The maser source, IRAS 19312+1950, exhibits various peculiarities. Its extended infrared nebulosity was first noticed in the 2MASS image by \citet{nak00}. The angular size of the near-infrared structure is roughly 30$''$. Soon after the initial identification of extended structure in the 2MASS image, a near-infrared (NIR) image with high angular resolution was obtained by the SIRIUS camera mounted on the University of Hawaii 2.2~m telescope \citep[see, Figure~1 in][]{deg04}. This image showed a point-symmetric structure with complicated filaments in its nebulosity. Since SiO maser emission has been detected toward IRAS~19312+1950 \citep{nak00}, the largely extended NIR structure is truly remarkable. This is because SiO maser sources, which are usually identified with an oxygen-rich (O-rich) asymptotic giant branch (AGB) star, rarely show extended structure that can be resolved by ground-based telescopes.

\citet{nak04b} discuss the spectral energy distribution (SED) of IRAS~19312+1950, which exhibits a flat-topped (or moderately double-peaked) profile (in a range from $\lambda=1$~$\mu$m to 100~$\mu$m) with a weak absorption feature of silicate at 9.7~$\mu$m.  They estimated a distance to the object by matching it to the absolute flux for a typical AGB star as 2.5--3.9 kpc [for reference, more recently, a trigonometric parallax distance to IRAS~19312+1950 was measured with VERA to be 3.8$^{+0.8}_{-0.6}$~kpc \citep{ima11}; this is roughly consistent with the distance given by Nakashima et al.~2004; in this paper we adopt 3.5~kpc as a representative distance if not otherwise specified]. This luminosity distance is not inconsistent with the kinematic distance estimated by assuming circular rotation for the Milky Way. However, a flat-topped SED is not usual in AGB stars, even though it is occasionally found in post-AGB stars \citep[see, e.g.,][]{uet00} and young stellar objects (YSOs).

\citet{mur07} analyzed an $H$-band image taken with the CIAO camera mounted on the 8~m Subaru telescope.  They constructed a radiative transfer model for the dust envelope by assuming a spherical geometry, to derive some of the fundamental stellar parameters for the central star; these are broadly consistent with an evolved stellar status. Further, \citet{nak04b} have estimated its mass-loss rate at $\sim10^{-4}$~M$_{\odot}$~year$^{-1}$ from their CO observations; this large mass-loss rate suggests that if the central star is an evolved star, it must be at a late-AGB (or early post-AGB) phase, where a dense superwind can take place \citep[in fact, such a large mass-loss rate has been identified in relatively young post-AGB stars; see, e.g.,][]{hri05}.

IRAS~19312+1950 exhibits a pretty complicated chemistry. So far, 22 molecular species have been detected \citep{deg04,nak04b}. The profile of the molecular radio lines consists of two different kinematic components: broad-weak ($\Delta v \sim 30$ km~s$^{-1}$) and narrow-strong ($\Delta v < 5$ km~s$^{-1}$) components. The peak velocities of these two components are roughly same. \citet{nak04b} found that the broad-weak component is emitted from a relatively small region with a angular size less than $\sim15''$ and the structure of the narrow-strong component is clearly resolved even by a single-dish telescope beam. 

\citet{nak05} made interferometric observations using BIMA in the CO ($J=1$--0 and 2--1), $^{13}$CO ($J=1$--0 and 2--1), C$^{18}$O ($J=1$--0), CS ($J=2$--1), SO ($J_{K}=3_{2}$--$2_{1}$), and HCO$^{+}$ ($J=3$--2) lines, and the structure of both the broad-weak and narrow-strong components were spatially resolved. The BIMA observation revealed that the morphological and kinematic properties of the broad-weak component are explained by an expanding sphere, while those of the narrow-strong component are best interpreted as a bipolar flow. The angular size of the narrow-strong component is 5-6 times larger than that of the broad-weak component. This larger size of the narrow component further complicates the situation, because a system consisting of a smaller spherical wind and larger bipolar flow is not naturally explained within the standard scheme of the stellar evolution unless the bipolar flow had a faster velocity than the spherical component at some point in the past.

One possible interpretation of IRAS~19312+1950 might be an O-rich AGB or a very young O-rich post-AGB star contingently embedded in an isolated small dark cloud \citep{deg04}. However, at the current moment, we have to say that we cannot be definitive about the true nature of this star. In this paper, we report the result of the first VLBI observations of IRAS~19312+1950 in the SiO and H$_2$O maser lines using the Very Long Baseline Array\footnote{The VLBA(HSA)/National Radio Astronomy Observatory is a facility of the National Science Foundation, operated under a cooperative agreement by Associated Universities, Inc.} (VLBA), High Sensitivity Array (HSA) and Japanese VLBI Network (JVN)\footnote{NRO and VERA/Mizusawa VLBI observatory are branches of the National Astronomical Observatory, an interuniversity research institute operated by the Ministry of Education, Culture, Sports, Science and Technology.}. We also report the result of radio interferometric observation of the OH maser line using the Multi-Element Radio Linked Interferometer Network (MERLIN). The purpose of this research is to aid our understanding of the true nature of IRAS 19312+1950 through the maser properties revealed by the interferometric observations. The outline of this paper is as follows. In section 2, we give details of the observation and data reduction. In section 3, we present the results of observations. In section 4, we discuss the kinematic properties of maser emission regions, and consider the evolutionary status of IRAS~19312+1950 on the basis of the present observations. Finally, the results are summarized in Section 5.


\section{Observations and Data Reduction}
\subsection{HSA Observation of the SiO Maser Lines}
The HSA observation of the SiO $J=1$--0 $v=1$ and SiO $J=1$--0 $v=2$ lines toward IRAS~19312+1950 was made on 2005 May 1, from 09:00 to 13:00 UT. The array used 10 VLBA antennas, 22 VLA phased-up antennas, and the Green Bank Telescope (GBT). The $v=1$ and $v=2$ lines were recorded separately in two different IF channels. Both left- and right-hand polarization signals were recorded for each IF channel simultaneously. The bandwidth of each IF channel was 8~MHz, corresponding to a velocity range of 56.0 km~s$^{-1}$. The central velocity of the IF channels was set to coincide with a systemic velocity at $V_{\rm lsr}=39.8$ km~s$^{-1}$. Due to hardware limitations, the VLA backend could only observed the $v=1$ line. Each IF channel consisted of 128 frequency channels, and the velocity resolution was therefore 0.43 km~s$^{-1}$. The coordinate value of the target was determined by the fringe-rate mapping after the clock parameter correction as: $\alpha_{J2000}=$19$^{h}$33$^{m}$24$^{s}$$\hspace{-2pt}.2452$$\pm$0$^{s}$\hspace{-2pt}.0035, 
$\delta_{J2000}=$~$+$19$^{\circ}$56$^{\prime}$55\arcsec.689$\pm$0\arcsec.075. 

We used the SiO $J=1$--0 $v=2$ line for fringe-fitting, because the $v=2$ line was much brighter than the $v=1$ line. We applied the calibration solutions obtained using the SiO $J=1$--0 $v=2$ line to all the spectral channels, using one reference channel at the intensity peak of the SiO $J=1$--0 $v=2$ line. Since the frequency separation between the $v=1$ and 2 lines is large, the phase delay is not properly calibrated at the frequency of the $J=1$--0 $v=1$ line, and therefore systematic uncertainty is expected in the offset coordinates. The degree of this positional uncertainty is theoretically estimated to be $\sigma \approx \sigma_{\alpha, \delta} \Delta_{\nu} / \nu \sim$0.6~milliarcsecond (mas); here, $\sigma_{\alpha, \delta}$ is the coordinate uncertainty of IRAS~19312+1950 ($\sigma_{\alpha,\delta}\sim$90~mas), the frequency separation between the two SiO maser lines is $\Delta\nu=$0.302~GHz, and the rest frequency of the SiO $J=1$--0 $v=2$ line is $\nu=$42.850~GHz. The amount of the angular shift seen in the map suggests that the maser spots of the $v=1$ and 2 lines are located in the same position within our estimated uncertainty. [Note that our definition of the "maser spot" is a point source (or a partially resolved small source) of maser emission, which is seen in each velocity channel.] We used AIPS for calibration, and then the image processing (including forming image cubes from visibilities, performing CLEAN, etc.) was carried out with Miriad \citep{sau95}. Robust weighting of the visibility data, which is an optimized compromise between natural and uniform weighting, gave a 0.48~mas~$\times$ 0.20~mas CLEAN beam with a position angle of $1.1^{\circ}$. The image cubes were created with a pixel size of 0.05~mas over 1024$\times$1024 pixels and 128 frequency channels.

\subsection{VLBA Observation of the H$_2$O Maser Line}
The VLBA observations of the H$_2$O maser line at 22.235~GHz ($6_{1,6}$--$5_{2,3}$) were made on 2004 August 23 and December 5 (project codes BI30A and BI30B). The observation used a part of a track for checking a system setup, and therefore the on-source integration time was limited in a short time ($\sim$17~min per day; the main purpose of the track was to do the astrometry of W43A, and we looked at IRAS~19312+1950 in a short time as a confirmation of the detection of the H$_2$O maser line). We used 9 VLBA antennas, and the received signals were recorded on two IF channels. The bandwidth of each IF channel was 8~MHz, corresponding to 54 km~s$^{-1}$ in velocity. The recorded signals were correlated with the Socorro FX correlator. Each IF channel was correlated to produce 512 frequency channels, and the velocity resolution was therefore 0.21~km~s$^{-1}$. We observed in dual polarization mode. The blue- and red-shifted components of the H$_2$O maser line (mentioned later in Section 3.1) were covered by different IF channels (we correlated RR and LL). Basic calibrations were carried out using AIPS through a standard procedure, and image processing (including forming image cubes from visibilities, performing CLEAN, etc.) was performed using Miriad. The bright quasar J2148$+$0657 was used for the instrumental delay calibration. The velocity channel at $V_{\rm lsr}=17.3$~km~s$^{-1}$ was used as a reference for the fringe-fitting and self-calibration. The calibration solutions obtained at $V_{\rm lsr}=17.3$~km~s$^{-1}$ were applied to all other channels. The robust weighting of the visibility data gave a 1.21~mas~$\times$~0.43~mas CLEAN beam with a position angle of $-51^{\circ}$. The image cubes were created with a pixel size of 0.12~mas over a grid of 1024$\times$1024 pixels and 512 frequency channels. The 1$\sigma$ noise levels in the maps are 19 mJy~beam$^{-1}$ and 29~mJy beam$^{-1}$ for the data of BI30A and BI30B, respectively.

\subsection{JVN Archival Data of the H$_2$O Maser Line}
We used archival data from the JVN for comparison with the VLBA data of the H$_2$O maser line. JVN consists of 6 antennas: four 20~m telescopes of the VLBI Exploration of Radio Astrometry (VERA) instrument, the 45~m telescope at the Nobeyama Radio Observatory, and the 34~m telescope operated by the National Institute of Communication Technology \citep[see, e.g.,][for the technical details about JVN]{ima06}. The H$_2$O maser line of IRAS~19312+1950 was observed with JVN on 2004 May 22 (Project code: r04143a). The total observation time was about 10 hours. The received signals were recorded in two IF channels in only left-had circular polarization with a bandwidth of 16~MHz each corresponding to a velocity coverage of 216~km~s$^{-1}$. The recorded data were correlated with the FX correlator of the National Astronomical Observatory of Japan (NAOJ). Each IF channel was correlated to produce 1024 channels, which give a velocity resolution of 0.21~km~s$^{-1}$. A single circular polarization pair (LL) was obtained. We used a bright maser spot at $V_{\rm lsr}=16.65$~km~s$^{-1}$ for fringe-fitting and self-calibration. Natural weighted visibility gave a synthesized beam of 1.27~mas~$\times$~0.82~mas with P.A.$=-46^{\circ}$. The data cube has a pixel size of 0.1~mas over a 1024$\times$1024 grid by about 500 frequency channels, and the 1$\sigma$ level in the map is 25~mJy~beam$^{-1}$.

\subsection{MERLIN Observations of the OH and H$_2$O Maser Lines}
The MERLIN observations in the three OH maser lines (1612, 1665, and 1667 MHz) and the H$_2$O maser line (22.235 GHz) were made on 2003 April 29--30 and 2002 December 19, respectively. The allocated telescope time was 13 hours each for the OH and H$_2$O observations (three OH lines were observed simultaneously). We periodically observed a bright quasar J1935$+$2031 as a phase calibrator to tack the phase variation over the time. We used 3C~84 for passband calibration. The cycle of the observational sequence was 24 minutes both for the OH and H$_2$O observations. In a single 24-minute sequence, we observed for 14 minutes on IRAS~19312+1950, for 7 minutes on the phase calibrator, for 3 minutes on the passband calibrator (More precisely speaking, in a single 24-miniutes sequence, we repeated a short sequence 7 times; each short sequence including a 2 minutes integration on IRAS~19312+1950 and 1 minute integration on the phase calibrator). Each on-source integration time includes a slewing time of the telescope (slewing times are roughly 15 seconds [between the target and phase calibrator] and 60 seconds [between the target and passband calibrator] ). The OH and H$_2$O lines were recorded in individual IF channels with bandwidths of 256~kHz and 2~MHz, respectively. The bandwidths of 256~kHz and 2~MHz correspond to velocity ranges of 45 km~s$^{-1}$ and 27 km~s$^{-1}$ respectively. Both polarization correlation pairs (RR and LL) were obtained. Only the blue-shifted component of the H$_2$O maser line at $V_{\rm lsr}=15$--20 km~s$^{-1}$ was covered in the MERLIN observation. Each IF channel was correlated to produce 128 frequency channels, with resulting nominal velocity resolutions of 0.35 km~s$^{-1}$ and 0.21 km~s$^{-1}$ respectively. After a-priori calibrations for visibility amplitude scales, basic visibility calibration of the raw data were carried out using AIPS, and image processing (including forming image cubes from visibilities, performing CLEAN, etc.) was performed using Miriad. Robust weighting of the visibility data yielded a 0.18$''$~$\times$~0.16$''$ CLEAN beam with a position angle of 20.3$^{\circ}$ for the OH observations, and a 24~mas~$\times$~15~mas CLEAN beam with a position angle of $-$49.6$^{\circ}$ for the H$_2$O observations. Image cubes of the OH and H$_2$O data were created with pexel sizes of 20~mas and 5~mas, respectively. We applied phase-referencing in the MERLIN observations, so that we could obtain the absolute coordinates of the H$_2$O maser source. 
The coordinate values obtained by Gaussian fitting are
 ${\rm R.A.}=$19$^{h}$33$^{m}$24$^{s}$\hspace{-2pt}.355$\pm$0.002,
${\rm Dec.}=$~$+$19$^{\circ}$56$^{\prime}$55\arcsec.701$\pm$0.02 (J2000). 
The OH maser main lines at 1665~MHz and 1667~MHz were not detected at the 3$\sigma$ level. The 1$\sigma$ levels at 1665~MHz and 1667~MHz were 2.5~mJy~beam$^{-1}$ and 2.8~mJy~beam$^{-1}$, respectively.


\section{Results}
\subsection{Observational Properties of the H$_{2}$O Line}
The top panel in Figure~1 shows the VLBA total-intensity line profile of the H$_{2}$O maser line together with the MERLIN profile. Both blue- and red-shifted components are clearly detected at $\sim$17~km~s$^{-1}$ and $\sim$51~km~s$^{-1}$ by the VLBA; MERLIN only observed the blue-shifted component. The peak velocities of the blue- and red-shifted components are consistent with the previous single-dish observations \citep{nak00,nak07}. Archival data from the Effelsberg/Medicina Monitoring Program \citep[see, e.g.,][]{eng97,win08} show a possible third peak at $V_{\rm lsr}\sim0$~km~s$^{-1}$, but unfortunately this component was not covered in the present observations. The VLBA and MERLIN observations detected almost the same source brightness. This fact suggests that the VLBA observation have not lost a significant amount of flux density due to spatial filtering. However, we have to bear in mind that the circumstellar masers usually exhibits a time variation in intensity \citep[see, e.g.,][]{win08}. The line parameters (i.e., the radial velocity, peak intensity and velocity integrated intensity) are summarized in Table~1. 

The emission regions of the red- and blue-shifted components are spatially separated with an angular distance of approximately 10.9~mas between the intensity peaks of the two components, as seen in the top and middle panels of Figure~2. The angular separation corresponds to a linear separation of 38~AU ($5.7\times10^{14}$~cm) at the adopted source distance of 3.5~kpc. The projected position angle of the line passing through the two intensity peaks of the blue and red-shifted components on the plane of the sky is about 108$^{\circ}$ (measured north to west). This position angle is somewhat different from that of the molecular bipolar flow mapped by BIMA \citep[$\sim$130$^{\circ}$;][]{nak05} and infrared extended structure \citep[$\sim$143$^{\circ}$;][]{mur07}. In the lower panel of Figure~1, we present the position-velocity diagram of the H$_2$O line. The emission feature seems to be a bipolar outflow rather than a Keplerian rotation disk (we discuss details about this matter in Section 4.1).

We compared the VLBA data with the archival JVN data to determine the relative proper motions of the maser spots. The three epochs spanned a 6 month period (the observing dates were 2004 May 22, 2004 Aug 23 and 2004 Dec 5). The results of the comparison is presented in Figure~2 (top panel). In order to compare the maser spot distribution, the maps of the different epochs were mutually aligned with respect to a position-reference spot at $\sim16$~km~s$^{-1}$. The reference spot survived over the observing period. Even though the number of detected maser spots varied from epoch to epoch, no clear proper motion was found (except for one spot; see, the last paragraph of Sect 4.1) over this 6 month period. The upper limit of the proper motion is 0.5~mas~yr$^{-1}$. 

To allow a better understanding of the kinematics of the blue-shifted component, we present the velocity channels maps and $p$--$v$ diagram in Figure~3. The blue-shifted component exhibits a velocity gradient in the $p$--$v$ diagram (see, upper-left panel in Figure~3; the gradient is roughly $-1.0$~mas/km s$^{-1}$), and in the channel maps we found in fact that the intensity peak position is shifting to the southeast direction as the velocity increases. The position angle of this projected motion on the plane of the sky is about 120$^{\circ}$; this direction is roughly coincident with that of a straight line passing through the two intensity peaks of the blue- and red-shifted components.

\subsection{Observational Properties of the SiO Maser Line}
Figure~4 shows the total intensity line profiles of the SiO $J=1$--0 $v=1$ and 2 lines superimposed on a single-dish spectrum taken by the Nobeyama 45~m telescope \citep{nak07}. Line parameters are summarized in Table~1. The intensity of the $v=2$ line is larger than that of the $v=1$ line, as confirmed in the previous single-dish observations. Such a relatively large intensity in the $v=2$ line is often seen in AGB stars with a low color-temperature \citep{nak03}. The red-shifted peaks are clearly detected at $v_{\rm lsr}\sim52$~km~s$^{-1}$ in both the $v=1$ and 2 lines, although the peak intensities are slightly smaller than those of the single-dish data. The red-shifted component of the $v=2$ line seems to have a weak wing-components peaked at 48.1 and 56.5~km~s$^{-1}$ with a S/N ratio of about 3. The intensity-peak positions of these weak components are almost exactly at the map center (see, corresponding channels in Figure~5). 

The blue-shifted component is not detected in the $v=1$ line, but in the profile of the $v=2$ line a faint possible detection of this component is found. We applied natural weighting for imaging this component to optimize the sensitivity (while we applied robust weighting for other components). Note that the line profile of the blue-shifted component given in Figure~4 was enlarged  40 times in intensity. The velocity separation between the blue- and red-shifted component is 36.3~km~s$^{-1}$ (from peak to peak), this separation is similar to that of the H$_2$O line (33.4~km~s$^{-1}$).

In previous single-dish observations, we have often simultaneously detected both the blue- and red-shifted components, but the intensity is time variable \citep[and occasionally the relative intensity of the two components inverts;][]{nak00,deg04,nak07}. (Note: the double peaks have been detectable in previous observations both in the $v=1$ and $v=2$ lines. However, the situation is a little bit different between the two transitions. In the case of the $v=2$ line, since the line intensity is relatively strong, we can almost always detect two peaks at a time. On the contrary, in the case of the $v=1$ line, usually only the stronger peak can be detected due to its relatively week intensity, but two peaks surely exist in the v=1 line.) Therefore, the blue-shifted components might be in a low-intensity phase in the present observations. Alternatively, the flux might be resolved out by interferometry. We produced a map consisting only of visibility data from short baselines ($<6\times10^{5}$~k$\lambda$), but the intensity of the line was too weak for a meaningful analysis. At the current moment, we have to say that the detection of the blue-shifted component of the $v=2$ line is tentative. Interestingly, however, the relative location of the possible blue-shifted component (with respect to the red-shifted component) is almost exactly the same with as of the H$_2$O maser line. The relative positions of the blue-shifted components (with respect to the red-shifted component) of the SiO and H$_2$O lines are $(\Delta{X}, \Delta{Y})=(-11\;{\rm mas}, -3\;{\rm mas})$ and $(\Delta{X}, \Delta{Y})=(-10\;{\rm mas}, -3\;{\rm mas})$, respectively.

The bottom panels in Figure~4 shows the total intensity maps of the red-shifted components in the $v=1$ and 2 lines (here we omit the map of the blue-shifted component, because the structure is not resolved). Both lines show an elongation in the northeast to southwest direction (this is presumably not the same elongation in the point spreading function), and we can see intensity peaks in the northeast and southwest lobes (particularly noticeable in the $v=2$ map). The angular separation between the two intensity peaks in the northeast and southwest lobes are different between the $v=1$ and $v=2$ lines: i.e., 0.72~mas and 1.3~mas in the $v=1$ and $v=2$ lines [these corresponds to $3.8\times10^{13}$~cm and $6.8\times10^{13}$~cm AU at 3.5~kpc]. The position of the central intensity peak of the $v=1$ line is shifted about 0.1 mas to the south and 0.4~mas to the west from the offset origin (i.e., (0, 0) position). This shift is likely instrumental however, because as stated in Section 2.2 we have calibrated the phase variation using the $v=2$ line that is distant from the $v=1$ line in frequency. In Figures 5 and 6, we present the channel velocity maps and position velocity diagrams. The northeast lobe has a slightly smaller systemic velocity than that of the central emission at $\sim51.8$~km~s$^{-1}$, while the southwest lobe has almost the same velocity as the central emission.

\subsection{Observational Properties of the OH Maser Line}
The upper-left panel in Figure~7 shows the total-intensity profile of the OH maser satellite line (i.e., the 1612~MHz transition), superimposed on a single-dish spectrum (1612, 1665 and 1667~MHz) taken by the Arecibo telescope (B. M. Lewis 2000, private communication). Compared to the Arecibo spectrum of the 1612~MHz line, the majority of the single-dish flux density seems to be recovered in the present interferometry. Since the MERLIN spectrum of the 1612~MHz line shows almost the same profile as the Arecibo spectrum, the line profile of the 1612~MHz OH maser line seems to be stable over the time scale of several years. The line profile is unusual for a proposed AGB star, i.e., a single, strong intensity peak is seen at $\sim29$~km~s$^{-1}$, and weak spikes are lying at lower velocities of the primary peak. This profile is clearly different from a typical case in AGB stars, which show a double-peak profile caused by the approaching and receding sides of an expanding spherical envelope. Here, we note that no emission is found at a velocity exceeding 50~km~s$^{-1}$, even though the Arecibo observation covered the velocity up to 70~km~s$^{-1}$.

The lower-left panel of Figure~7 shows the total-flux intensity map of the OH 1612~MHz line. The feature is partially resolved by the present synthesized beam. The $p$--$v$ diagram given in the right panels in Figure~7 seems to exhibit a systematic motion in the velocity range lower than 27~km~s$^{-1}$: in the upper panel of the $p$--$v$ diagram the emission peaks at velocities less than 27~km~s$^{-1}$ are moving away from the offset origin as the velocity increases, while the emission peaks in the lower panel remain stably on the offset origin over the entire velocity range. This systematic motion is also confirmed in the velocity-channel maps in Figure~8, in which the emission peak is moving away from the map center to the eastward direction as the velocity increases (but the emission peak of the strongest emission at 28.0--31.6~km~s$^{-1}$ is at the map center).


\section{Discussion}
The present observations in maser lines have revealed the kinematics of maser emission regions in the molecular envelope of IRAS~19312+1950 for the first time. As mentioned in Section~1, one of the most important scientific problems on this peculiar IRAS object is the identification of its evolutionary status. In this section, we first discuss the kinematic properties of the H$_2$O and SiO maser emission region, and consider how the OH maser properties can be understood within in the context of the stellar evolution. Then, we revisit the discussion about the evolutionary status of IRAS~19312+1950 on the basis of the present maser observations and previous observational results.

\subsection{Properties of the H$_2$O and SiO Masers and its Kinematic Interpretation}
The most notable results in the present observations are visible in Figure~2. The features seen in the top and middle panels of Figure~2 are clearly different from the maser-spot distribution of the H$_2$O maser of typical AGB stars. Since the usual H$_2$O maser emission of AGB stars predominantly comes from a spherically expanding envelope, the distribution of maser spots are usually much more extended or scattered, and often exhibits a ring (or patchy ring) structure \citep[see, e.g.,][]{bow93,bow94}. In the case of IRAS~19312+1950, the detected maser spots are lying roughly on a straight line as seen in the top panel of Figure~2. This maser spot distribution is quite different from that of typical AGB stars. Even though the existence of aspherical motions are occasionally suggested in AGB stars, based on VLBI observations in the H$_2$O and SiO maser lines \citep[see, e.g.,][]{dia03,nkg08}, even in such cases a spherical component is still dominant, and therefore the maser spot distribution is much more scattered compared with the case of IRAS~19312+1950.

Then, how can we interpret the kinematics of the maser emission region? One possibility might be a Keplerian rotating disk showing a double-peak line profile, which originates in the two edges of the disk. If we assume that the distance to the object is 2.5--3.9~kpc \citep{nak04b} and Keplerian rotation, the central mass is estimated to be 4.4--6.7 M$_{\odot}$. This is not inconsistent with AGB mass (i.e., 0.8--8.0 M$_{\odot}$), and the existence of a Keplerian rotation disk is sometimes suggested in AGB/post-AGB stars and PNe \citep[see., e.g.,][]{ber00,cot04,buj05,nak05,bab06,der07}. 

However, Keplerian rotation disks in AGB stars present significant challenges in angular momentum transport, absent a binary companion. For IRAS~19312+1950, firstly the feature seen in the $p$-$v$ diagram  (see, lower panel of Figure~1) does not show a Keplerian-rotation-like curve, which should show a point-symmetric pair of two arches. Moreover, if we assume a rotating disk, we should see emission at velocities close to the systemic velocity, because the material lying in the near- and far sides of the disk has no relative velocity to us, but no such emission is seen in the spectra. (The spectra given in Figure~1 have a gap at velocities close to the systemic velocity, but we have confirmed that there is no emission in this velocity range through previous single-dish observations.) Additionally, we also should see emission at the spatially-intermediate location between the blue and red-shifted components if we assume a rotating disk model \citep[see, e.g.,][]{bab06}, but no such emission is confirmed in Figure~2. For these reasons, it seems that a Keplerian rotation disk is most likely excluded.

Another possibility might be a binary system consisting of two maser sources. The tentative detection of the SiO maser (mentioned in Section 3.2) suggests this idea, because the location of the tentative SiO emission (i.e., blue-shifted component), which coincides with the H$_2$O maser emission, is somewhat unusual for a single AGB star. Maser spots of the SiO, H$_2$O and OH maser lines usually exhibit a stratified distribution in an AGB envelope in order of decreasing excitation temperature (i.e., SiO masers locate in the innermost region, H$_2$O masers are surrounding the SiO maser emission, and OH maser locate in the outermost region). Therefore if the tentative detection of the SiO maser emission is real we need to consider the reason for the spatial coincidence, although we do note however that the relative location of different maser species can be affected by projection effects if there is a highly asymmetric distribution of circumstellar material.

One possible explanation for the spatial coincidence of different masers and the kinematic properties of the H$_2$O emission region might therefore be a binary system including two maser stars (AGB stars or red supergiants, for example) emitting both H$_2$O and SiO masers. This is most likely not the case however, because the maser spot distribution again presents a problem: the straight-line distribution of maser spots seen in Figure~2 is not consistent with the maser spot distribution of AGB stars and red supergiants, which show more scattered distribution originated in a spherical envelope. Even if we assume that the straight-line distribution can be explained, for example, by tidal interaction, the total luminosity of the binary system including two maser stars raises further problems. Two AGB stars increase the absolute flux of the binary system up to at least 16,000 -- 20,000 L$_{\odot}$ (if we assume that red supergiants are included in the system, the total luminosity must be more than that; see, Section 4.3), and this large absolute flux causes an unreasonably large distance, which does not match up with the other physical parameter such as the kinematic and trigonometric parallax distances and infrared flux. Therefore the explanation of a binary system of two maser stars is ruled out here. 

If we try to understand the spatial coincidence of the SiO and H$_2$O masers (if this is real), we should consider the other possibilities: for example, a peculiar pumping mechanism different from that working in usual circumstellar envelopes of AGB stars. However, before going to such an astrophysical interpretation of SiO maser emission, we should confirm the detection of the blue-shifted component of the SiO maser line by doing further VLBI observations, because the detection is just tentative at this time.

For the moment, a reasonable explanation of the kinematics of the maser emission region would be a bipolar outflow. Even though we have not very clearly confirmed a proper motion of maser spots in a 197-days period, this is presumably because the observing period is too short and/or the effects of the inclination angle. In Section 3.1, we mentioned the velocity gradient ($\sim$1.0 mas/km~s$^{-1}$) found in the blue-shifted component of the H$_2$O line. If we assume a Hubble-type flow (i.e., the motion in which the velocity increases in proportional to the distance from the central star), we can estimate the distance (and dynamical timescale) between the driving source of the jet and the blue-shifted component. However, the estimated distance, 18~mas, does not match up with the geometrical center of the maser spot distribution as seen in Figure~2. Presumably, the kinematics of the jet is not a simple Hubble-type flow, and the acceleration seems to change intricately with the time and distance. 

On the other hand, if we assume a ballistic motion of the jet and find a proper motion of maser spots, we can estimate the dynamical timescale. With the same method applied in our previous studies (see, e.g., Imai et al.~2002, Yung et al.~2010), as mentioned in Sect 3.1 we found a proper motion of only one maser-spot, which exhibits a proper motion of 0.64 mas along with the jet axis (see, the arrow in Figure 2). The dynamical time scale of the jet based on this proper motion (assuming a ballistic motion) is about 5.5 years. However, we should note that this value of a proper motion is not very reliable, because we seen a possible acceleration in Figure~3, and therefore because the identification of the maser spots over different epochs is not very reliable. To reveal the details of the jet kinematics, we need to monitor the proper motion of maser spots in a much longer time scale.

\subsection{Properties of the OH Maser}
According to a previous single-dish observation (B. M. Lewis 2000, private communication), the 1612~MHz line is clearly stronger than the 1665 and 1667~MHz lines, and therefore IRAS~19312+1950 should be classified into the type~II OH maser source \citep[][see, Figure~7: Note that in Figure~7 the Arecibo profiles of the 1665 and 1667~MHz lines are magnified in intensity]{hab96}. OH maser emission is often detected in YSOs, but the 1665 and 1667~MHz lines are always stronger than the 1612~MHz line in YSOs \citep[see., e.g.,][]{edr07}, and therefore the relative intensity ratio of the OH maser lines supports the evolved star status of IRAS~19312+1950.

As well as the H$_2$O and SiO masers, however, the line profile of the 1612~MHz OH maser is also clearly different from the typical case of AGB, which usually show a double-peak profile \citep{dia85,siv90,dav93} caused by the approaching and receding sides of an expanding spherical envelope. Note that AGB and post-AGB stars occasionally show a single-peak profile, but such a single-peak profile usually can be interpreted as a part of a double-peak profile originated from a spherically expanding shell. In the case of IRAS~19312+1950, emission peaks of the 1612~MHz OH line are seen only in the blue-shifted side of the systemic velocity ($\sim$34--36~km~s$^{-1}$), and the line profile is very complicated, consisting of a single strong peak and multiple weak spikes as seen in Figure~7. 

Since \citet{nak05} detected a compact spherical flow (with a size of $~$3--5$''$) in their CO line mapping, the distribution of the OH maser spots might trace the inner part of the compact spherical flow. In fact, if we assume that we are looking at a part of a spherically expanding shell with the systemic velocity of $\sim$34--36~km~s$^{-1}$, the behavior of OH maser spots seen in the $p$--$v$ diagram (in PA$=$90$^{\circ}$) might be explained by a spherically expanding shell with an expanding velocity of 15--25~km~s$^{-1}$ \citep[in fact, the compact spherical flow detected in the CO lines has an expanding velocity of roughly 20~km~s$^{-1}$,][]{nak05}. Here we note that the size of an spherically expanding shell should be growing up as the velocity comes closer to the systemic velocity. However, a problem is that such a spherically-expanding-shell-like feature cannot be found at PA$=$0$^{\circ}$, suggesting the prospective spherical flow somehow has deficiency of a part of its structure. This asymmetric nature might be explained by the effect of a bipolar flow lying at the central part of the envelope, otherwise it might be explained by a torus-like structure (having its axis roughly in the direction of PA$=$90$^{\circ}$), which blanks out the spherical flow. Unfortunately, in the present case, it is quite difficult to check the sphericity of the emission region of the 1612 MHz line, because the angular resolution of the MERLIN observation is insufficient. Interestingly, the Arecibo profiles of the 1665 and 1667~MHz lines exhibit an intensity peak shifted from that of the 1612 ~MHz line. Since the OH main lines (i.e., 1665 and 1667~MHz lines) are excited closer to the central star than the 1612~MHz line in general \citep[see, e.g.,][]{cha94}, this velocity shift might suggest the systematic difference of the gas motions between the outer and inner regions. 

\citet{dea04} investigated the relation between mid-infrared colors and the line profile of the OH maser lines, and found that there is a likely evolutionary trend in the shape of the OH maser profile with some sources evolving from double-peaked to irregular (caused by a bipolar jet) profiles. The red mid-infrared color of IRAS~19312+1950 [$\log(F_{25}/F_{12})=0.5$; here, $F_{12}$ and $F_{25}$ are the IRAS flux densities at $\lambda=25$ and 12~$\mu$m, respectively] suggests that this object is lying at the very late stage of AGB or already at early post-AGB \citep{nak00}, and therefore it is not very strange even if an irregular profile and asymmetric nature caused by the interaction between a spherical AGB wind and a bipolar flow is seen in the OH maser properties, because well-developed bipolar jets are often found in post-AGB stars and planetary nebulae.

\subsection{Considerations on the Evolutionary Status}

The detection of the SiO maser emission seems to support the evolved star status of IRAS~19312+1905, because SiO maser sources are usually identified to evolved stars except for three, anomalous YSOs in star forming regions \citep[i.e., Ori IRc 2, W51 IRs 2, and Sgr B2 MD5; see, e.g.,][]{has86}. On the other hand, the largely extended nebulosity, rich-chemistry and large mass of IRAS~19312+1950 \citep{nak04b,deg04} have caused a problem in the identification of its evolutionary status. Even though we have repeatedly discussed this problem in our previous papers \citep{nak00,nak04a,nak04b,deg04,nak05,mur07}, the current data provide a fresh insight into this question. (For readers' convenience, in Table 2 we briefly summarize the discussions about the evolutional status in previous papers.)

As we discussed in Sections 4.1 and 4.2, the kinematics of the maser emission region most likely can be explained as a bipolar outflow, but molecular bipolar flows are seen in a wide variety of objects: YSOs, AGB and post-AGB stars, proto-planetary nebulae (PPNe; note that some groups apply the word ``pre-planetary nebulae'' to PPNe to avoid confusion with proto-planetary disks in star forming regions) and planetary nebulae (PNe), and therefore the bipolarity of the molecular outflow, by itself, cannot be a definitive probe of the evolutionary status. However, the other properties of the maser emission seem to support the evolved star hypothesis, as we mentioned above: for example, the intensity ratios of the OH maser lines (i.e., classification into the type II OH maser sources) are consistent with the evolved star status, and the kinematical nature of the OH maser is reminiscent of an spherically expanding flow (although some asymmetric nature is found). Additionally, we note that there are no known OH maser sources toward low-mass YSOs \citep[see, e.g.,][]{gar99,edr07,sah07}, and therefore if IRAS~19312+1950 is a YSO, it must be related to a high-mass star forming region (SFR), suggesting the presence of an ultracompact H$_{\rm II}$ region with detectable continuum emission. However, the NRAO VLA Sky Survey at $\lambda=$20~cm \citep{con98} shows no continuum sources toward IRAS~19312+1950. In addition, Br$\gamma$ emission was also negative in a several-minute integration using the 2.3~m telescope at the Australian National University (P. R. Wood 2002, private communication). Even though there are three exceptional SiO maser sources in star SFRs \citep{nak00,nak05} as mentioned above, all the three sources exhibit a strong continuum emission of an ionized region, because those are lying in high-mass SFRs. Thus, we can say that the nature of IRAS 19312+1950 is clearly different from that of the SiO maser sources previously known in SFRs.

If we assume IRAS 19312+1950 is an evolved star enshrouded by a cold envelope, what kind of evolved stars are appropriate for the true identity of this object according to its observational properties? Although red supergiants (RSGs) often emit SiO masers, we can exclude RSGs from the candidates, because its large luminosity (e.g., $>2\times10^{4}$~L$_{\odot}$, Wood et al.~1983; $>1.3\times10^{5}$~L$_{\odot}$, Groenewegen et al.~2009) causes an unreasonably large distance, which does not match up with other physical parameters such as the kinematic distance \citep{nak04b} and trigonometric parallax distance \citep{ima11}. Therefore, we should consider only intermediate- or low-mass evolved stars that have a main-sequence mass less than about 8.0~$M_{\odot}$. Although intermediate-mass stars finally reach to the PPN and PN phases via the AGB and post-AGB phases, PNe and PPNe harboring an ionized region also can be excluded from the possibilities, because no continuum emission is observed indicating an ionized region. Practically, possibilities are limited to AGB and post-AGB stars, which do not harbor a significant ionized region but can  emit SiO maser emission.

In our previous papers \citep{nak00,deg03,deg04,nak04a,nak04b,nak05}, we have mentioned the resemblance between IRAS~19312+1950 and OH~231.8+4.2 (Rotten Egg Nebula), which is a PPN with SiO maser emission. In fact, OH~231.8+4.2 exhibits a rich set of molecular species \citep{mor87}, SiO and H$_2$O maser emission \citep[see, e.g.,][]{des07} and an extended bipolar nebulosity \citep{mea03}. The infrared colors of OH~231.8+4.2 is also similar to that of IRAS~19312+1950 \citep[see, e.g.,][]{nak00}. The H$_2$O maser emission coming from a tip of a bipolar jet was mapped with VLBA \citep{des07}, showing a similar morphological and kinematic properties to IRAS~19312+1950. However, the nature of the SiO maser emission seems to be different in the two objects. In the case of OH~231.8+4.2, a M9-10 III Mira variable is lying at the center of the nebulosity, and the circumstellar rotating ring (or disk) around this Mira variable is considered to be the source of the SiO maser emission \citep{des07}. On the contrary, so far we have no observational evidence of a Mira variable in IRAS~19312+1950; for example, a monitoring observation in near-infrared band for roughly one-month period did not show any evidence of a time variation in intensity (Fujii et al. 2002, private communication) and the SiO maser properties does not show a rotation or spherical expansion, which is a prospective characteristic of circumstellar SiO maser emission of a Mira-type star. At the current moment, we believe that the nature of the SiO maser emission of IRAS~19312+1950 is different from that of OH~231.8+4.2.

Here, we point out notable resemblance between IRAS 19312+1950 and a particular class of young post-AGB (or late-AGB) stars, based on the maser properties: namely the ``water fountains (WFs)'' -- that is young post-AGB (or late-AGB) stars exhibiting a velocity range of the H$_2$O maser emission larger than that of the OH maser emission \citep[see, e.g.,][]{ima02}. Since water fountains are harboring a high velocity, tiny molecular jet, which is usually detected in the H$_2$O maser line, the H$_2$O line shows a large velocity range. Tiny molecular jets found in WFs have been considered to be the onset of the formation of asymmetric PN morphology. Therefore, the kinematics of the tiny molecular jets in WFs has been frequently investigated using the VLBI technique \citep[see, e.g.,][]{ima02,vle06,yun10}. The primary characteristics of WF jets are its high expanding velocity (often larger that 100 km~s$^{-1}$) and highly collimated morphology, and in addition the jets occasionally exhibit a precessing/spiral pattern \citep{ima02,yun10}. OH maser emission is detectable from the mojority of know WFs \citep[a dozen of WFs are known so far; see, Table 1 in][]{ima07} except for IRAS 19134+2131 (for this object, no OH emission is detected; the WF status is confirmed only with the high velocity range of the H$_2$O maser emission). Since the nature of OH maser emission of WFs are basically same with that of AGB stars \citep[a typical expansion velocity of the OH line is 10--25 km~s$^{-1}$; see, e.g.,][]{tel89}, the emission seems to come from an spherical envelope, which is not yet affected by the bipolar jet \citep[a spherical distribution of OH maser emission is confirmed with VLBI observations; see, e.g.,][]{ima02}.

In the case of IRAS~19312+1950, according to Figure~7, the expansion velocity of the OH component is roughly 15--25~km~s$^{-1}$ if we assume the systemic velocity of the object is 35~km~s$^{-1}$ (this expansion velocity is consistent with that of other known WFs). On the other hand, if we take into account the 0~km~s$^{-1}$ component (detected in previous observations; see, Section 3.1), the expansion velocity of the H$_2$O component must be larger than 25~km~s$^{-1}$. This situation is quite reminiscent of WFs, even though in the case of IRAS~19312+1950 only the blue-shifted side of the OH emission is detected. In fact, infrared colors of IRAS~19312+1950 are very similar to those of W43A, which is a proto-typical example of a water fountain \citep[see Figure~6 in][]{nak03}. There, as well as in IRAS~19312+1950, SiO maser emission, which exhibits a double peak profile, is detected, and the velocity separation of the two peaks is roughly 30~km~s$^{-1}$ \citep{nak03,ima05}; this SiO maser properties of W43A is similar to those of IRAS~19312+1950. Additionally, the large mass-loss rate estimated from CO observations \citep[$\sim10^{-4}$~M$_{\odot}$~yr$^{-1}$;][]{nak04b} is also consistent with the WF status, because the mass-loss rate of intermediate-mass AGB stars can often reach such a high mass-loss rate at late-AGB or early post-AGB due to the super wind \citep[see, e.g.,][]{hri05}. What is more, we see the difference of position angles of bipolar structure between different observational methods, as mentioned in Section 3.1. This fact might be due to a precessing motion, which is often seen in water fountains \citep[see, e.g.,][]{ima02}.

The origin of the rich chemistry and large ambient mass of IRAS 19312+1950 is, however, still left unsolved. In order to consider the relation and resemblance between IRAS 19312+1950 and WFs, further investigations of WFs are required (for example, CO mapping, molecular line surveys and searching for new WFs), because the number of well-studied WFs are still very limited. Additionally, checking the sphericity of the emission region of the OH maser is also important; for this purpose, more sensitive (to observe 1665 and 1667 MHz lines) and higher-resolution (to resolve each maser emission source) VLBI observations of the OH maser lines are required.

\subsection{Short Remarks on Another Possibility}
As discussed in Section 4.3 and previous papers, in the standard scheme of stellar evolution, some observational properties of IRAS~19312+1950 cannot be explained unless we assume a rare situation (for example, a intermediate-mass evolved star embedded in a dark cloud). An interesting object related to this problem is the peculiar nova, V838~Mon. This object exhibits detectable SiO maser emission \citep{deg05}, and does not fall into any conventional categories of SiO maser sources. V838 Mon erupted in the beginning of 2002 January, exhibiting the spectrum of A--F-type supergiant around optical maximum, and then the spectral type has changed to that of an M-type supergiant \citep[therefore, bright in infrared bands,][]{mun02,cra03}. At the current moment, the merger of two stars, on the main sequence or evolving toward the main sequence, is touted as the most promising model to explain the observational properties of V838 Mon \citep{sok07,mas10}. This merging hypothesis might be an advantage when we explain the large ambient mass of IRAS~19312+1950, if the merged star exhibits a luminosity similar to a single AGB star (as a large luminosity causes a problem as discussed in Sections 4.1 and 4.3). In fact, \citet{kam08} observed the CO gas of V838 Mon, and the derived mass of the ambient material is roughly 25~M$_{\odot}$; this value is similar to the mass of IRAS~19312+1950 \citep[$\sim10$~M$_{\odot}$,][]{nak04b}. However, to trigger the merging event producing the V838 Mon-type object \citep[recently, the word of ``red novae'' has come into use for meaning this type of star, see, e.g.,][]{mas10}, the star should be lying in a star forming region or open cluster, in which main-sequence or young stars are tightly-packed. As we discussed in the previous papers, so far there is no clear evidence of star-forming activities around IRAS~19312+1950, but the census of the nearby stars might be useful to consider the possibility of the red nova origin of IRAS~19312+1950.


\section{Summary}
In this paper, we have reported the results of interferometric observations of IRAS 19312+1950 in the SiO, H$_2$O and OH maser lines. On the basis of maser properties revealed, we have considered the evolutionary status of this object. The main results are as follows:

\begin{enumerate}
\item The double-peak profile of the H$_2$O maser line is clearly detected by the present VLBA observation, and the emission regions of the red- and blue-shifted components are spatially separated at an approximate angular separation of 11 mas. No proper motion of maser spots is confirmed by comparing VLBA and JVN data at different epochs, but the kinematic properties of the maser emission region seem to be consistent with a bipolar flow rather than alternative interpretations such as a Keplerian rotating disk.
\item The red-shifted component of the SiO maser lines ($J=1$--0, $v=1$ and 2) is clearly detected. The blue-shifted components of the $J=1$--0, $v=2$ line is tentatively detected, but is very faint and the further confirmation is required. Interestingly, however, the intensity-peak separation of the red- and blue-shifted components coincides almost exactly with those of the H$_2$O line.
\item The 1612~MHz OH maser line exhibits a complicated line profile: a single strong peak plus high-velocity weak spikes. All the emission is detected on the blue-shifted side of the systemic velocity. The structure is partially resolved, and the kinematics of the maser emission region is reminiscent of a compact spherical flow, which has been found by the previous CO observation.
\item The maser properties appeared to be consistent with the evolved star hypothesis. In particular, we found a resemblance between IRAS 19312+1950 and water fountain maser sources, which are post-AGB stars harboring a high-velocity, bipolar molecular jet, even though the origin of some observational properties (mass and chemistry) are still unsolved.
\end{enumerate}


\acknowledgments
This work is supported by a grant awarded to JN from the Research Grants Council of Hong Kong (project code: HKU 703308P; HKU 704209P; HKU 704710P) and the Seed Funding Program for Basic Research of the University of Hong Kong (project code: 200802159006). HI and SD have been financially supported by Grant-in-Aid for Scientific Research from Japan Society for Promotion Science (18740109 and 20540234, respectively). The authors thank Drs. Tom Muxlow, Anita Richards, and Peter Thomasson for supporting the observation and data reduction. The NRAO's VLBA (and HSA) is a facility of the National Science Foundation of the USA, operated under a cooperative agreement by Associated Universities, Inc. We acknowledge all staff members and students who have helped in array operation and in data correlation of the JVN in order to obtain the present archival data. JN thanks Yong Zhang, Chih-Hao Hsia, Bosco H. K. Yung and Thomas K. T. Fok for stimulating discussions.



\begin{deluxetable}{llrrrr}
\tablecolumns{5}
\tablewidth{0pc}
\tablecaption{Line parameters of detected lines}
\tablehead{
\colhead{Transition} & \colhead{Equipment} & \colhead{$V_{\rm peak}$} &  \colhead{$T_{\rm peak}$}  & \colhead{$S$} & \colhead{obs. date} \\
\colhead{} & \colhead{} & \colhead{(km s$^{-1}$)} & \colhead{(Jy)} & \colhead{(Jy km s$^{-1}$)} &  \colhead{(yymmdd)} }
\startdata
H$_2$O 6$_{1,6}$--5$_{2,3}$ (blue comp.) & VLBA & 17.2 & 7.8 & 15.0 & 041205 \\
H$_2$O 6$_{1,6}$--5$_{2,3}$ (blue comp.) & MERLIN & 16.8 & 10.4 & 16.2 & 021219 \\
H$_2$O 6$_{1,6}$--5$_{2,3}$ (red comp.) & VLBA & 50.6 & 0.9 & 1.3 & 041205 \\
SiO $J=1$--0 $v=1$ (red comp.)& VLBA(HSA) & 51.9 & 0.21 & 0.39 & 050501 \\
SiO $J=1$--0 $v=2$ (red comp.) & VLBA(HSA) & 52.1 & 0.75 & 1.61 & 050501 \\
SiO $J=1$--0 $v=2$ (blue comp.)$^{*}$ & VLBA(HSA) & 15.7 & 0.023 & 0.048 & 050501 \\
OH $^{2}\Pi_{3/2}$ $J=3$/2 $F=1$--2 & MERLIN & 28.9 & 0.768 & 3.72 & 030429 \\
\enddata
\tablenotetext{*}{Tentative detection (see Section 3.2).}
\end{deluxetable}

\clearpage

\begin{deluxetable}{lccl}
\tablecolumns{4}
\tablewidth{0pc}
\rotate
\tablecaption{Summary of observational results}
\tablehead{
\colhead{Observation} & \colhead{Evolved star} & \colhead{YSO} & \colhead{Comment and reference}}
\startdata
Distance & X &  & {\tiny A trigonometric parallax distance supports the luminosity distance of an AGB/post-AGB star$^{1}$.} \\
OH maser intensity ratios & X &  & {\tiny 1612 MHz line is the strongest$^{2}$.} \\
Spherically expanding shell & X &  & {\tiny A spherically expanding molecular shell is detected in the CO 1--0 line$^{3}$.} \\
Isolation from nearby clouds & X &  & {\tiny MIR images show the object is clearly isolated$^{4}$.} \\
Mass-loss rate & X & & {\tiny Dust and molecular moss-loss rates are consistent with AGB or post-AGB$^{4,8}$.} \\
$^{12}$CO/$^{13}$CO ratio & x &  & {\tiny A small $^{12}$CO/$^{13}$CO intensity ratio of radio CO lines supports evolved stars$^{3,4}$.} \\
Detection of SiO maser & X & x & {\tiny Only 3 detections in YSOs$^{5}$, while more than 2000 detections in evolved stars$^{6}$.} \\
OH maser line profile & X & X & {\tiny Irregular profile$^{2}$. Indicating a possible spherically expanding shell.} \\
Bipolar molecular flow & X & X & {\tiny Detected both in a small$^{2}$ and large scales$^{3}$.} \\
IR morphology & X & x & {\tiny Point-symmetric$^{7}$. Typical in evolved stars, but still possible in YSOs.} \\
IR colors and SED & X & x & {\tiny Flat top SED$^{4}$. The color is not typical in YSOs, but still possible.} \\
Complex chemistry & x & X & {\tiny Rare in O-rich evolved stars$^{7}$, but still possible.} \\
Detection of the methanol line &  & X & {\tiny No previous detections in evolved stars$^{7}$.} \\
\enddata
\tablenotetext{X}{supportive}
\tablenotetext{x}{weakly supportive, or not impossible}
\tablenotetext{1}{\citet{ima10}}
\tablenotetext{2}{present paper.}
\tablenotetext{3}{\citet{nak05}}
\tablenotetext{4}{\citet{nak04b}}
\tablenotetext{5}{see, section 4.3 in the present paper}
\tablenotetext{6}{\citet{deg07}}
\tablenotetext{7}{\citet{deg04}}
\tablenotetext{8}{\citet{mur07}}
\end{deluxetable}

\clearpage


\begin{figure}
\epsscale{.70}
\plotone{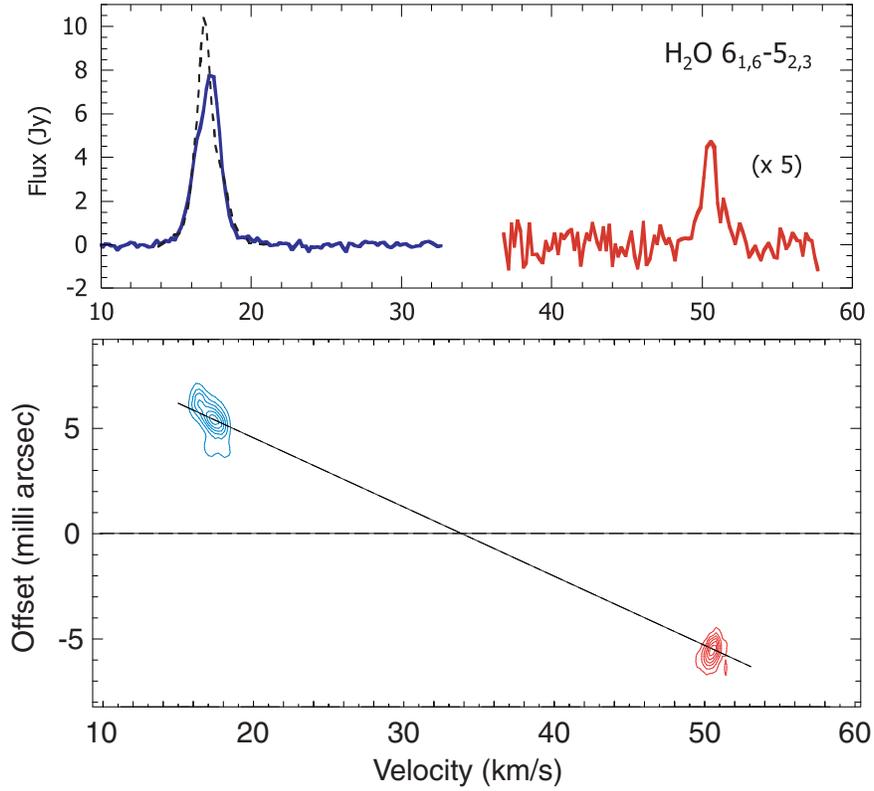}
\figcaption{{\it (Upper panel)}: Total flux profile of the H$_2$O maser line (6$_{1,6}$--5$_{2,3}$) taken by VLBA (red and blue lines) and MERLIN (black dotted line). {\it (Lower panel)}: Position-velocity diagram in the position angle of 107.5$^{\circ}$. The offset origin is taken at the middle point of the blue- and red-shifted components: (DEC offset, RA offset)$=$($-$5.72~mas, 1.88~mas). \label{fig1}}
\end{figure}
\clearpage

\begin{figure}
\epsscale{.60}
\plotone{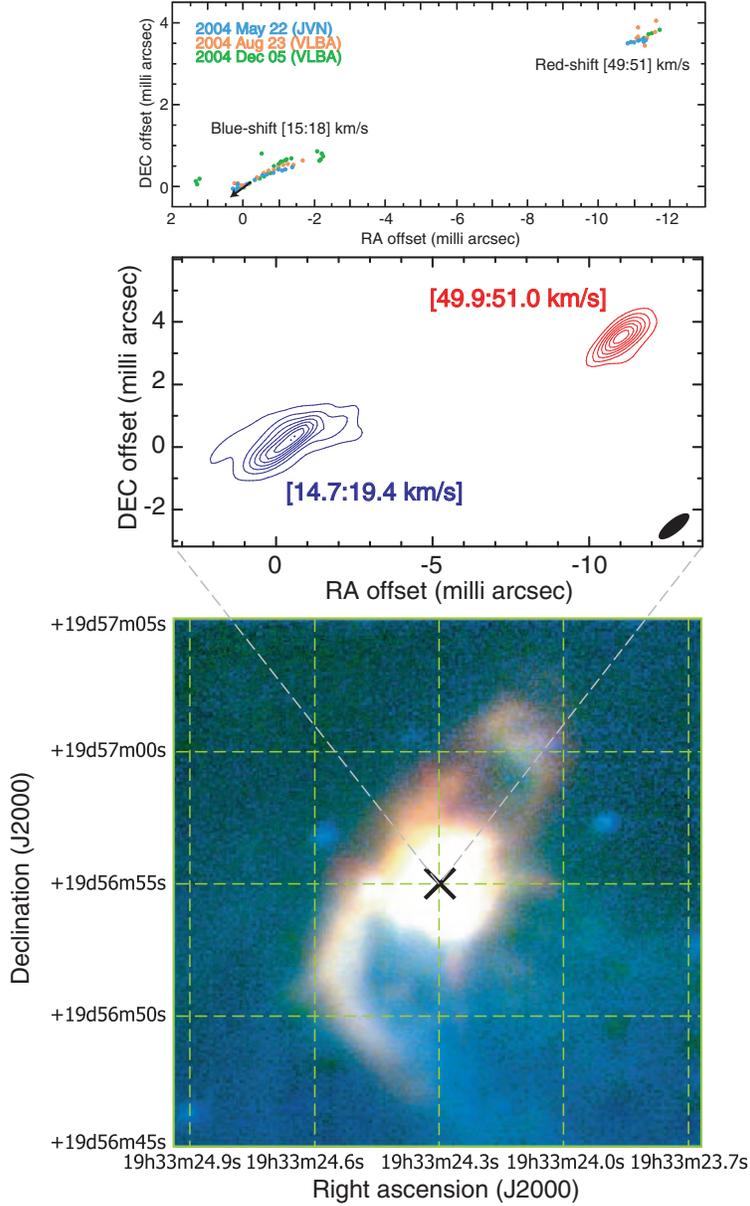}
\figcaption{{\it (Top)}: Spatial distribution of maser spots of the H$_2$O maser line (6$_{1,6}$--5$_{2,3}$). The blue, orange and green dots represent the JVN observation on 2004 May 22, VLBA observation on 2004 August 23, and VLBA observation on 2004 December 5, respectively. The black arrow represents the direction of the proper motion (see, the last paragraph in Sect 4.1). {\it (Middle)}: VLBA total intensity map of the H$_2$O maser line (epoch 2004 December 5). The contour levels of the blue-shifted component are 0.05, 0.10, 0.15, 0.19, 0.24, 0.29, 0.33, and 0.38~Jy~beam$^{-1}$. The contour levels of the red-shifted component are 0.04, 0.18, 0.32, 0.45, 0.59, 0.73, 0.86, and 1.00~Jy~beam$^{-1}$. The FWHM beam size is located in the bottom right corner. The velocities in the brackets represent the integral range. {\it (Bottom)}: Near-infrared H-band image taken by SUBARU/CIAO (Murakawa et al.~2007). \label{fig2}}
\end{figure}
\clearpage

\begin{figure}
\epsscale{.75}
\plotone{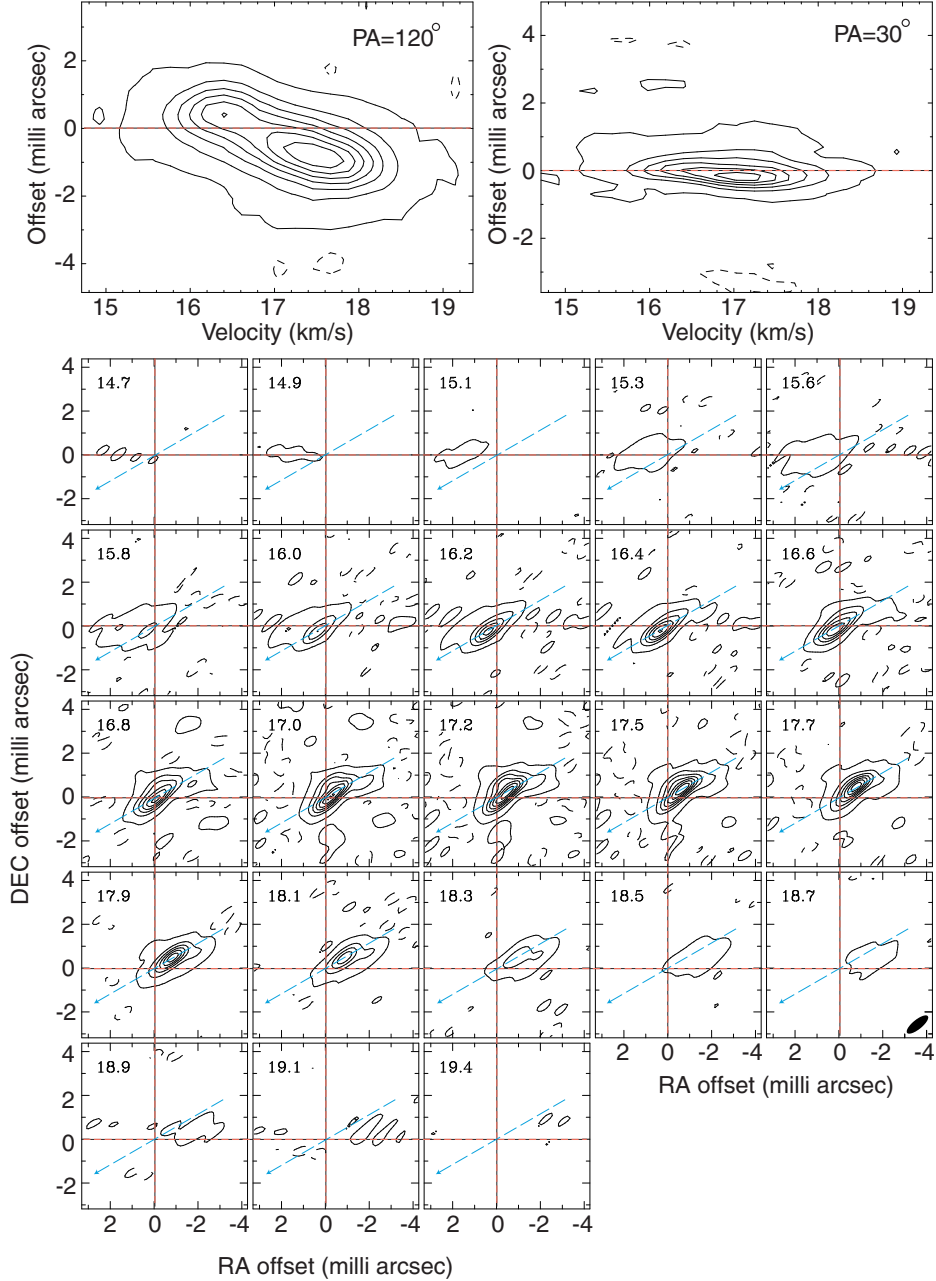}
\figcaption{{\it (Upper panels)}: Position-velocity diagram of the H$_2$O maser line (6$_{1,6}$--5$_{2,3}$) in the position angle of 120$^{\circ}$ and 30$^{\circ}$ [the PA$=$120$^{\circ}$ cut is indicated by the blue dotted arrows in the lower panel and the origin of the offset axis is taken at the phase center]. The contour levels are 0.06, 0.53, 1.00, 1.47, 1.94, 2.40, 2.87 and 3.34 Jy~beam$^{-1}$. {\it (Lower panels)}: Velocity--channel maps of the H$_2$O maser line. The contour levels are same with the top panels. The channel velocities are given in the upper left corner of each panel. The synthesized beam is indicated in the bottom right. \label{fig3}}
\end{figure}
\clearpage

\begin{figure}
\epsscale{.65}
\plotone{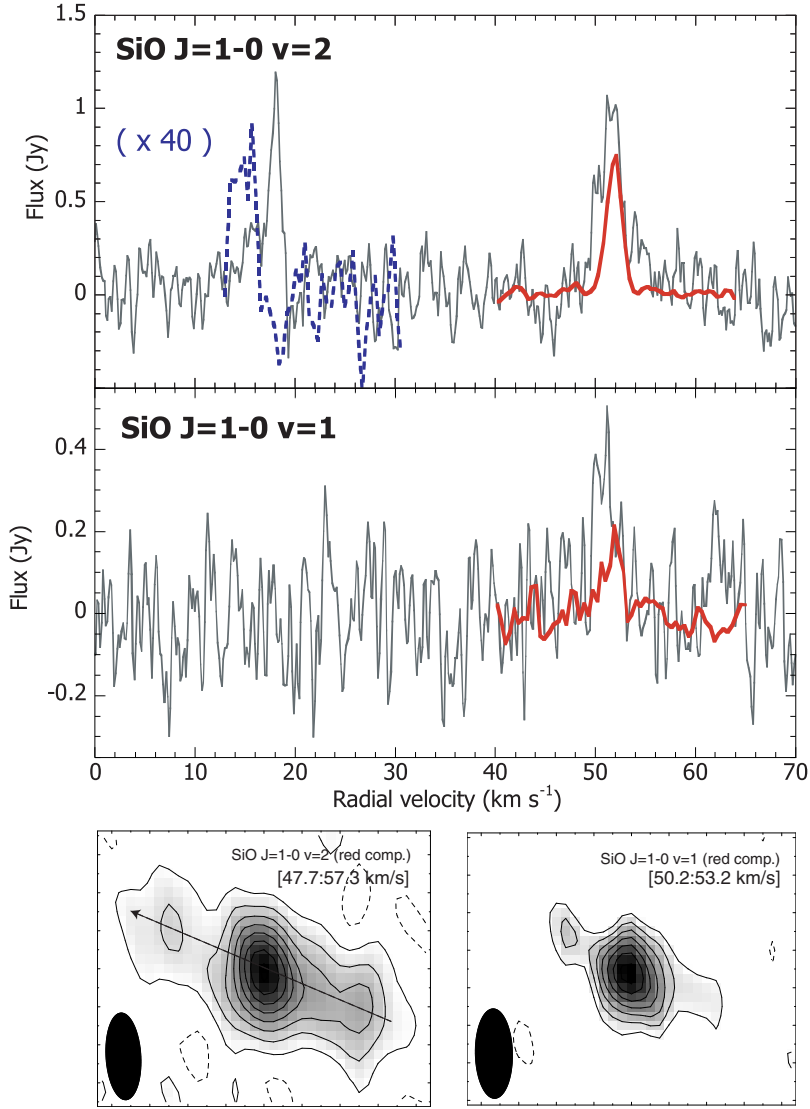}
\figcaption{{\it (Upper panels)}: VLBA total flux line profiles of the SiO $J=1$--0 $v=1$ and 2 lines (red and blue-dotted lines). The gray line represents the single dish profile taken by the Nobeyama 45~m telescope \citep{nak07}. The intensity of the blue-shifted component of the $J=1$--0 $v=2$ line is at a magnification of 40 times (note that the detection of the blue-shifted component is tentative; see text). {\it (Lower panels)}: VLBA total intensity maps of the red-shifted components of the SiO $J=1$--0 $v=1$ and 2 lines. The contours start from a 3~$\sigma$ level, and the levels in the $J=1$--0 $v=1$ and 2 maps are spaced every 1.42~$\sigma$ and 5.29~$\sigma$, respectively. 1~$\sigma$ levels in the $J=1$--0 $v=1$ and 2 maps are $4.3\times10^{-3}$~Jy~beam$^{-1}$ and $1.4\times10^{-3}$~Jy~beam$^{-1}$, respectively. The dashed contour corresponds to $-3$~$\sigma$. The FWHM beam size is located in the bottom left corners. The velocities in the brackets represent the integral range. The dashed arrow represents the cut (position angle of 67$^{\circ}$) used for Figure~6. \label{fig4}}
\end{figure}
\clearpage

\begin{figure}
\epsscale{.70}
\plotone{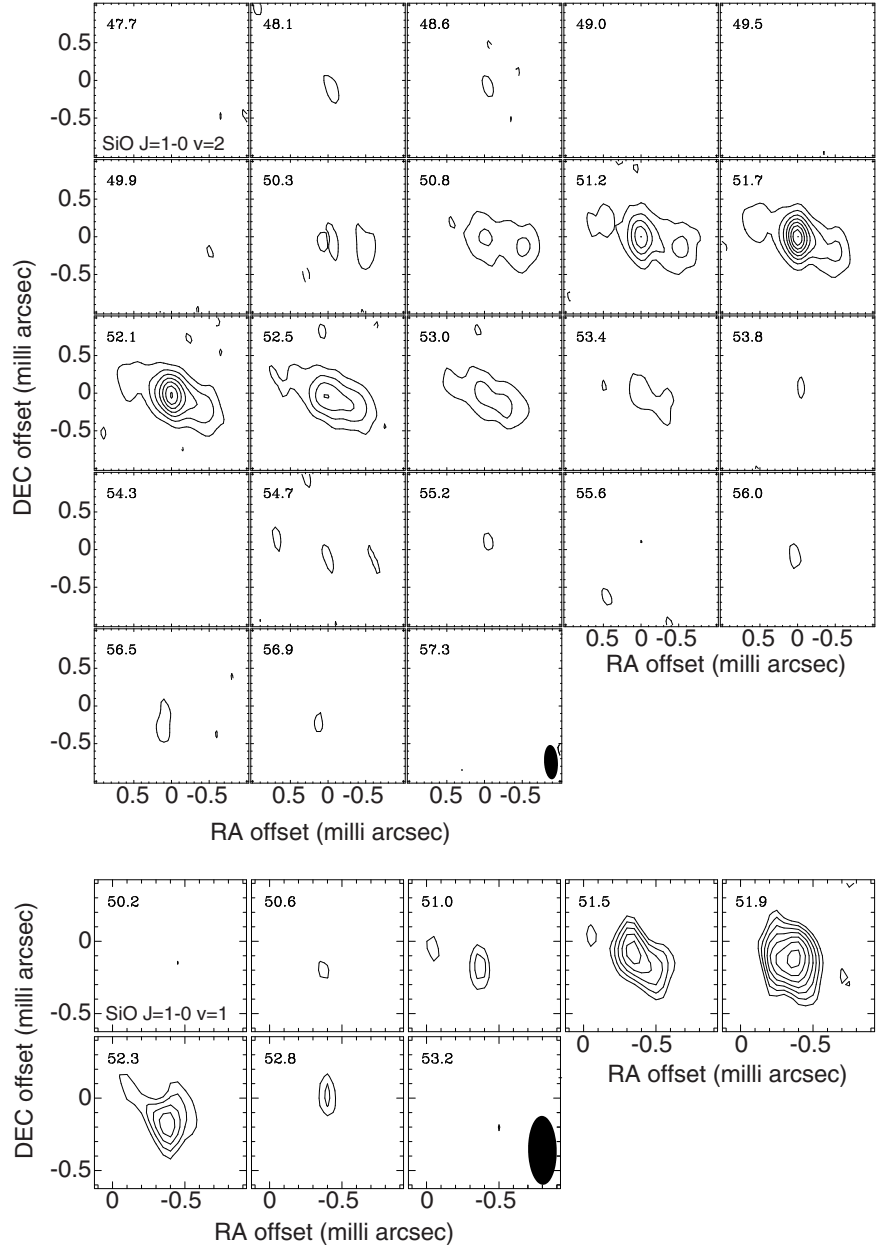}
\figcaption{{\it (Upper panels)}: Velocity channel maps of the SiO $J=1$--0 $v=2$ line. The contour levels start from a 3~$\sigma$ level, and the levels are spaced every 5.98~$\sigma$. 1~$\sigma$ corresponds to $6.9\times10^{-3}$~Jy~beam$^{-1}$. The dashed contour both in the upper and lower panels represents a $-3$~$\sigma$ level (even though almost no $-3$~$\sigma$ feature in the maps). The FWHM beam size is presented in the last channel. {\it (Lower panels)}: Velocity channel maps of the SiO $J=1$--0 $v=1$ line. The contour levels start from a 3~$\sigma$ level, and the levels are spaced every 0.83~$\sigma$. 1~$\sigma$ corresponds to $1.2\times10^{-2}$~Jy~beam$^{-1}$ The FWHM beam size is presented in the last channel. \label{fig5}}
\end{figure}
\clearpage

\begin{figure}
\epsscale{.40}
\plotone{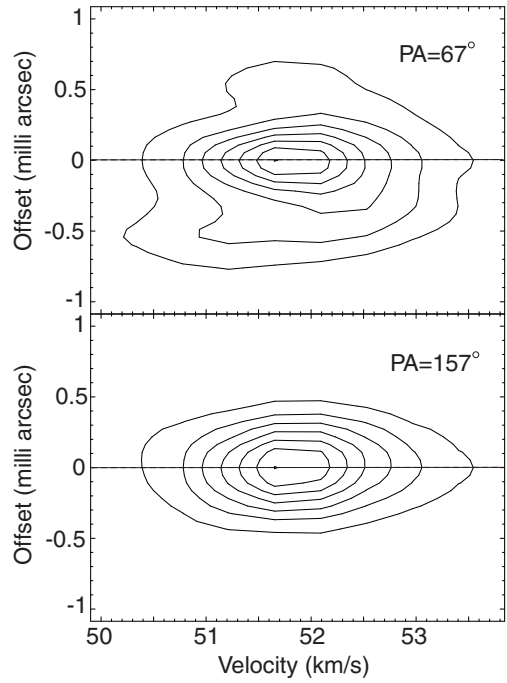}
\figcaption{Position-velocity diagrams of the SiO maser line ($J=1$--0 $v=2$) in the position angles of 67$^{\circ}$ and 157$^{\circ}$. The PA$=$67$^{\circ}$ cut is indicated in Figure~4 (see, the dashed arrow in the lower-left panels). The origin of the offset axis is taken at the phase center. The contour levels are 21, 47, 73, 99, 125, 151 and 177 mJy~beam$^{-1}$. \label{fig6}}
\end{figure}
\clearpage

\begin{figure}
\epsscale{.90}
\plotone{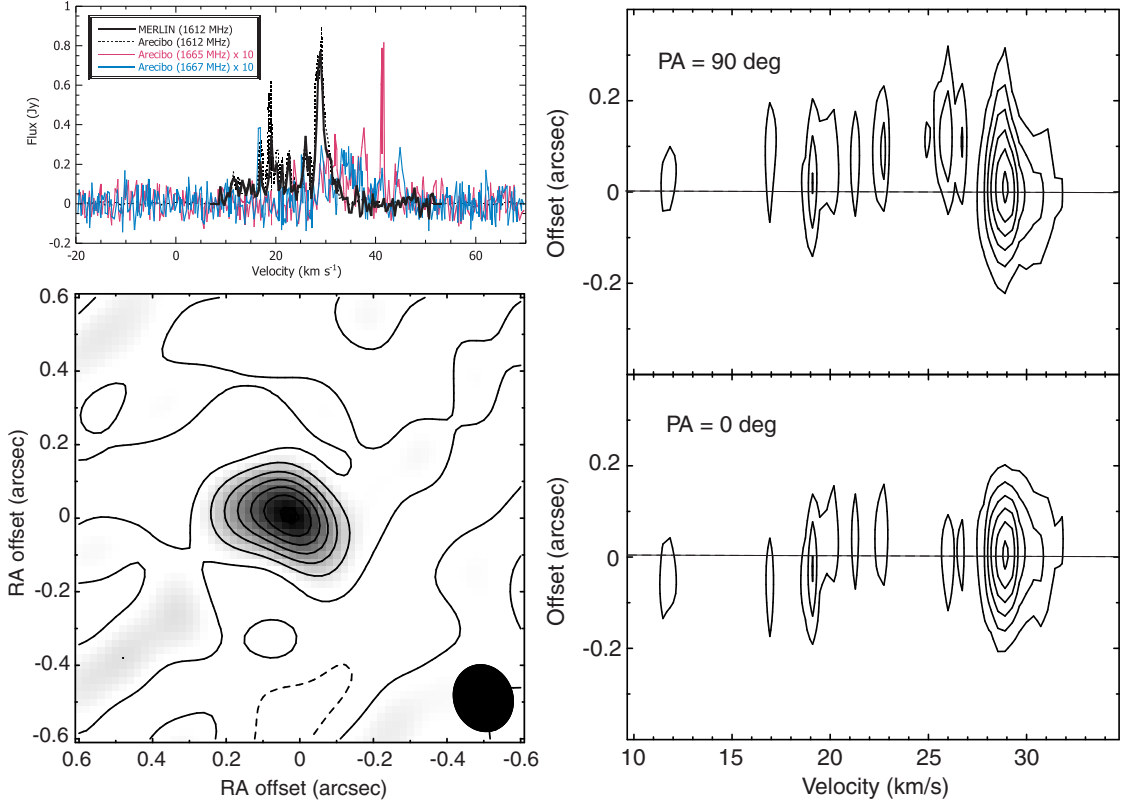}
\figcaption{{\it (Left-upper)}: Total flux profile of the OH maser satellite line (1612~MHz) taken by MERLIN (black solid line) superimposed on the profiles of the Arecibo data of the OH 1612, 1665 and 1667~MHz lines. {\it (Left-lower)}: MERLIN total intensity map of the OH maser line (1612~MHz). The contour levels are $6.5\times10^{-3}$, $2.3\times10^{-2}$, $3.9\times10^{-2}$, $5.5\times10^{-2}$, $7.1\times10^{-2}$, $8.7\times10^{-2}$ and $1.0\times10^{-1}$ Jy~beam$^{-1}$. The FWHM beam size is located in the bottom right corner. The dashed contour represents a $-3$~$\sigma$ level. {\it (Right panels)}: Position-velocity diagram of the OH maser line (1612~MHz) in the position angles of 90$^{\circ}$ and 0$^{\circ}$. The origin of the offset axis is taken at the phase center. The contour levels are 0.05, 0.12, 0.19, 0.27, 0.34, 0.41 and 0.48 Jy~beam$^{-1}$. \label{fig7}}
\end{figure}
\clearpage

\begin{figure}
\epsscale{.80}
\plotone{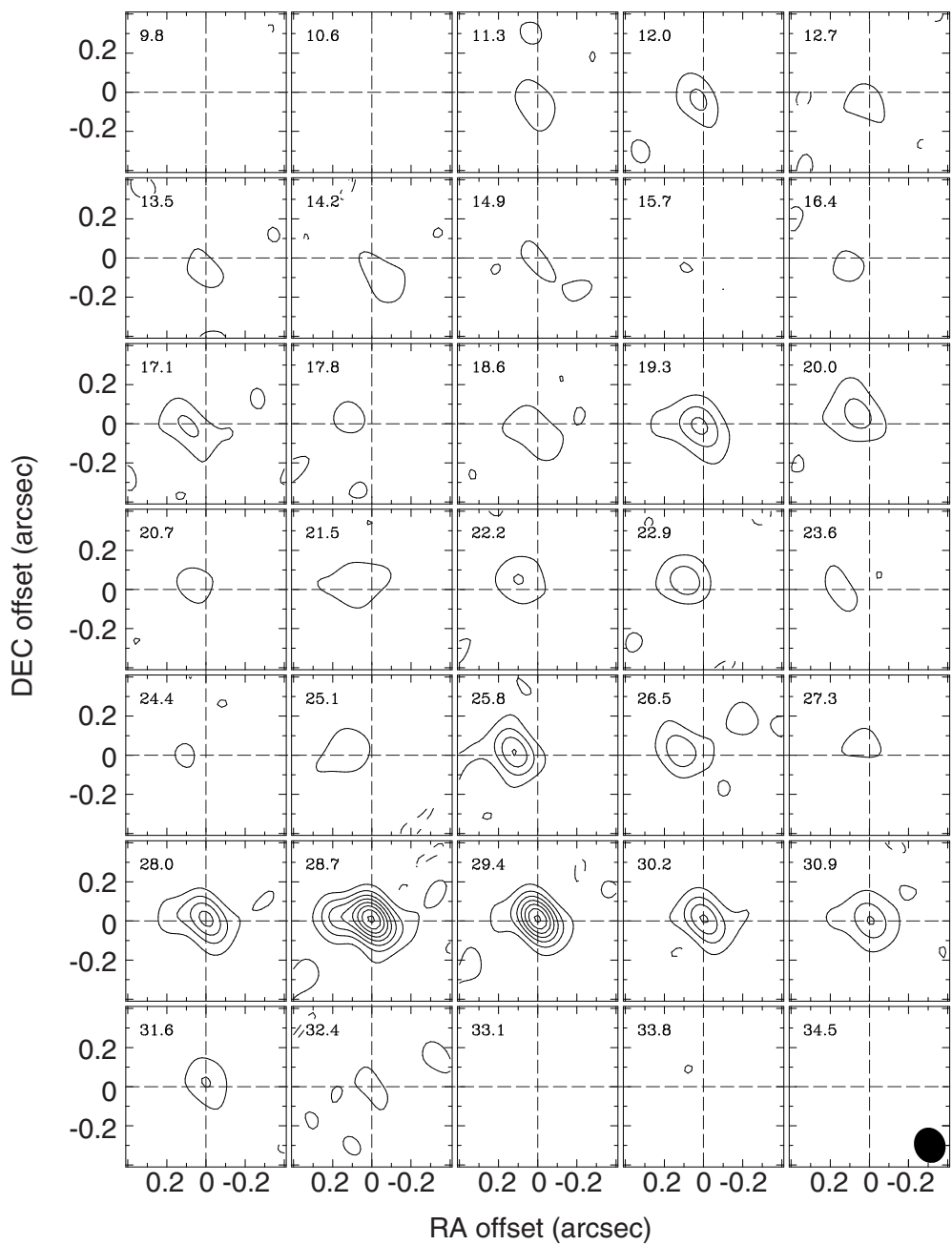}
\figcaption{Velocity--channel maps of the OH 1612~MHz line. The contours start from a 3~$\sigma$ level, and the levels are spaced every 6~$\sigma$. The 1~$\sigma$ level corresponds to $1.12\times10^{-2}$ Jy~beam$^{-1}$. The dashed contour corresponds to $-3~\sigma$. The peak intensity corresponds to 46~$\sigma$. The synthesized beam is indicated in the lower right corner. \label{fig8}}
\end{figure}
\clearpage

\end{document}